\documentclass[a4paper,11pt]{article}
\usepackage{jcappub}
\usepackage{graphicx}
\usepackage{pgfplots}
\usepackage{subfigure}
\usepackage{float}
\newcommand{\Lagr}{\mathcal{L}}
\pgfplotsset{height=5cm,width=0.5cm}

\title{Dynamics of a scalar field, with a double exponential potential, interacting with dark matter}

\author[a]{Vartika Gupta,}
\author[a]{Rakesh Kabir,}
\author[a]{Amitabha Mukherjee}
\author[a]{and Daksh Lohiya}

\affiliation[a]{Department of Physics and Astrophysics, University of Delhi, Delhi-110007, India.}

\emailAdd{varg@physics.du.ac.in}
\emailAdd{rakesh@physics.du.ac.in}
\emailAdd{am@physics.du.ac.in}
\emailAdd{dlohiya@physics.du.ac.in }

\abstract{We study the interaction between dark matter and dark energy, with dark energy described by a scalar field having a double exponential effective potential. We discover conditions under which such a scalar field driven solution is a late time attractor. We observe a realistic cosmological evolution which consists of sequential stages of dominance of radiation, matter and dark energy, respectively.}

\begin{document}
\maketitle
\flushbottom
\section{Introduction}
\ The {\it{``late time''}} accelerated expansion of the universe was discovered in 1998~\cite{perlmutter99}. A repulsive gravity-inducing (in other words, with negative pressure) {\it dark energy} component can cause such accelerated expansion. A cosmological constant,  having equation of state $w=\frac{p}{\rho}=-1$, is the simplest candidate for this dark energy~\cite{sahni00,sahni04}.
 
$\quad$ Although such a model is fairly concordant with observations, 
it suffers from two major issues: smallness of the
cosmological constant, and the coincidence problem. Its effects are becoming 
just noticeable in the recent history of the universe~\cite{copeland06}.
A way to alleviate these problems is to explore frameworks that cause an 
effective cosmological constant to dynamically arise out of interaction 
between dark 
energy and dark matter - thereby reproducing the late time accelerating behavior
quite akin to that produced by the cosmological constant~\cite{copeland06, boehmer08, copeland98}.

The present article describes dynamical features of tracker solutions that can 
be expected to arise from the dynamics of moduli fields that naturally occur 
in string theory~\cite{huey00}. These fields provide a 
parametrization of a compactified manifold, while their vacuum expectation 
values determine the effective four dimensional gauge and gravitational coupling
constants from the unification scale. It has been shown that in a universe 
dominated by other matter fields, the moduli fields can be dynamically stable. 
Away from the minimum, the evolution of the moduli fields can be expressed in 
terms of an effective scalar field having a double exponential(exponent of an exponential) potential 
 as considered in 
eq.~\eqref{potential}~\cite{ng01}.

In the next section, we introduce the notations and conventions. Further, keeping in sight the ensuing cosmological dynamics, we outline characteristic features of a
scalar field.   Thereafter cosmological evolution equations have been non-dimensionalized for the convenience in the dynamical analysis. Critical points for 
the resulting autonomous system and their stability issues are discussed in Sec. III, followed by results and conclusions in Sec. IV.

\section {Dynamics of canonical scalar field }
The action of a minimally coupled scalar field in a four dimensional spacetime 
 is given by
\begin{equation}
\label{eq:action}
S=\int d^4x\sqrt{-g}\;\big[\frac{R}{2\kappa^2}+ {\it \Lagr_\phi+\Lagr_m}\big] ,
\end{equation}
\\where {\it g } is the determinant of the metric, {\it R} is the Ricci scalar, 
$\kappa^2 = 8\pi G$ ($c=1$) , {\it $\Lagr_m$} \rm  is the matter Lagrangian, and the 
scalar field Lagrangian $\Lagr_\phi$ is given by
\begin{equation}
\label{eq:lagrangian}
{\it \Lagr_\phi}\rm=-\frac{\epsilon}{2} \partial_\mu \phi\partial^\mu \phi-V(\phi) 
\end{equation} 
\\with V being a general 
potential for $\phi$. Epsilon $\epsilon$ is equal to +1 for the quintessence, and $-1$ for the phantom field~\cite{chen09}. The variation of the metric gives the gravitational field equations:
\begin{equation}
\label{einstein}
G_{\mu\nu}=\kappa^2 (T_{\mu\nu[m]}+T_{\mu\nu[\phi]}) 
\end{equation}
\\where $G_{\mu\nu}\equiv R_{\mu\nu}-\frac{1}{2}g_{\mu\nu}R$ is the Einstein tensor. The 
matter stress-energy tensor for a fluid may be parametrized in terms of the  
four velocity of the fluid $u_\mu$, and the density and pressure functions 
 $\rho_m$ and $p_m$ as 
\begin{equation} 
T_{\mu\nu[m]}=(\rho_m+p_m)u_\mu u_\nu+p g_{\mu\nu} \,.
\end{equation} 
The scalar field energy-momentum tensor is
\begin{equation}
T_{\mu\nu[\phi]}=\epsilon\partial_\mu\phi \partial_\nu \phi+ g_{\mu\nu}\Lagr_\phi \label{5}\,.
\end{equation} 
\\\\We consider a spatially flat FRW metric with the metric tensor given by
\begin{equation}
g_{\mu\nu}=\rm diag (-1,\it a(t)^2,\it a(t)^2, \it a(t)^2)\rm  
\end{equation}
\\ where $a(t)$ is the scale factor. The scalar field is taken to be spatially 
homogeneous, i.e., $\phi=\phi(t)$. This gives
\begin{equation}
T_{0[\phi]}^0\equiv-\rho_\phi=-\frac{\epsilon}{2}\dot \phi^2-V(\phi)\,, 
 \end{equation}
 \begin{equation}
 \label{eq:space-energy-tensor}
  T_{1[\phi]}^1 = T_{2[\phi]}^2 = T_{3[\phi]}^3\equiv 
p_\phi=\frac{\epsilon}{2}\dot \phi^2-V(\phi) \,.
\end{equation}
\\ We assume that generic equation of state for the $\it{i}th$  component (radiation, dark matter or scalar field) is 
\begin{equation}\label{eos}
p_i=(\gamma-1)\rho_i 
\end{equation}
\\ where $\gamma$ is a constant. Gamma ($\gamma$) is equal to 4/3 for radiation, and 1 
for pressureless dark matter. 
For the scalar field, relation~\eqref{eos} gives  the effective equation of state as 
\begin{equation}
\gamma_\phi=\frac{p_\phi+\rho_\phi}{\rho_\phi}=\frac{2\dot \phi^2}{\dot \phi^2+2V} \,.
\end{equation}
\\Eq.~\eqref{einstein} relates the Hubble parameter $H=\dot a/a$ to the stress-energy tensor components to give the Friedmann constraint
\begin{equation}
\label{friedmann-constraint}
H^2=\frac{\kappa^2}{3}(\rho_m+\rho_\phi+\rho_{rad}) \,,
\end{equation}
\\ and the Raychaudhuri equation
\begin{equation}
\label{raychaudhuri}
\dot H=-\frac{\kappa^2}{2}(\rho_m+\frac{4}{3}\rho_{rad}+\epsilon \dot \phi^2)\,.
\end{equation}
\\Finally, the conservation equations for the field ($\phi$), 
cold dark matter(m) and radiation(rad) read
\begin{eqnarray}
\dot\rho_m+3H\rho_m=0 \label{matter-evolution}\,,\\
\dot\rho_\phi+3H\rho_\phi\gamma_\phi=0 \label{phi-evolution}\,, \\
\dot\rho_{rad}+4H\rho_{rad}=0 \label{rad-evolution} \,.
\end{eqnarray}

So far we have not considered any coupling between the dark matter and the scalar field. 
Introducing an effective coupling denoted by $Q_I$ between dark energy and dark matter modifies Eqs.~\eqref{matter-evolution} and~\eqref{phi-evolution} to 
\begin{eqnarray}
\dot \rho_m+3H\rho_m=Q_I \label{matter-evolution-int}\,, \\
\dot \rho_\phi+3H\rho_\phi\gamma_\phi=-Q_I \label{phi-evolution-int}\,
\end{eqnarray}
\\while the modified Klein-Gordon equation is given by
\begin{equation}\label{modified-KG}
\ddot \phi +3H\dot \phi +\frac{dV}{d\phi}=-\frac{Q_I}{\epsilon\dot \phi}. 
\end{equation}
\\Eqs.~\eqref{raychaudhuri},~\eqref{matter-evolution-int},~\eqref{phi-evolution-int} and ~\eqref{modified-KG} are the evolution equations subject to the Friedmann constraint~\eqref{friedmann-constraint}. Consider 
dimensionless variables~\cite{boehmer08,ng01,amendola00}: \\\\
\begin{equation}\label{def:var}
x\equiv \frac{\kappa\dot \phi}{\sqrt{6}H},\quad y\equiv \frac{\kappa\sqrt{V}}{\sqrt{3}H},\quad u\equiv\frac{\kappa\sqrt{\rho_{rad}}}{\sqrt{3}H},\quad z\equiv\frac{-1}{V\kappa}\frac{dV}{d\phi} \,.
\end{equation}\\
\\In terms of these new dimensionless variables, the evolution equations lead to a generalized 
four dimensional autonomous system \\\\
\begin{subequations}\label{dyn-sys}
\begin{align}
x^\prime &=-\frac{3x}{\epsilon}+\sqrt{\frac{3}{2}}y^2z+\frac{3}{2}x[(1-\epsilon x^2-y^2-u^2)+\frac{4}{3}u^2+2\epsilon x^2]-\frac{Q}{\epsilon} \,. \\
y^\prime &=-\sqrt{\frac{3}{2}}yxz+\frac{3}{2}y[(1-\epsilon x^2-y^2-u^2)+\frac{4}{3}u^2+2\epsilon x^2]\,. \\
u^\prime &=-2u+\frac{3}{2}u[(1-\epsilon x^2-y^2-u^2)+\frac{4}{3}u^2+2\epsilon x^2] \,. \\
z^\prime &=-\sqrt{6}z^2x(\Gamma-1) \,.
\end{align}
\end{subequations}
\\Here prime denotes differentiation with respect to the new time variable 
$N\equiv ln(a/a_{0})$ and we have further defined
\begin{equation}\label{def:third-var}
z\equiv -\frac{1}{\kappa V}\frac{dV}{d\phi}\,,
\end{equation}
\begin{equation}\label{def:gamma}
\Gamma \equiv V\frac{\frac{d^2V}{d\phi^2}}{(\frac{dV}{d\phi})^2} \,,
\end{equation}
\begin{equation}
 Q\equiv \frac{\kappa Q_I}{\sqrt{6}H^2\dot \phi} \,. \label{def:interaction}
\end{equation}
\\ As the dynamical system ~\eqref{dyn-sys} is invariant under 
$y\rightarrow -y$, we need to solve the system only for $y\geq 0$. 
\\The Friedmann constraint ~\eqref{friedmann-constraint} gives
\begin{equation}
\frac{\kappa^2\rho_m}{3H^2}+\epsilon x^2+y^2+u^2=1 
\end{equation}
\\or

\begin{equation}
 \Omega_\phi+\Omega_r+\Omega_m=1 
\end{equation}
 where
\begin{equation}
\Omega_\phi\equiv \frac{\kappa^2\rho_\phi}{3H^2},\qquad \Omega_r\equiv \frac{\kappa^2\rho_{rad}}{3H^2},\qquad \Omega_m\equiv \frac{\kappa^2\rho_m}{3H^2} \,.
\end{equation}
\\Also $\Omega_\phi$ is bounded, i.e., $ 0\leq \epsilon x^2+y^2\leq 1$ for non-negative $\Omega_r$ and $\Omega_m$.\\

\begin{figure}[t!]
\centering
\subfigure[]{\includegraphics[width=60mm, height=54mm]{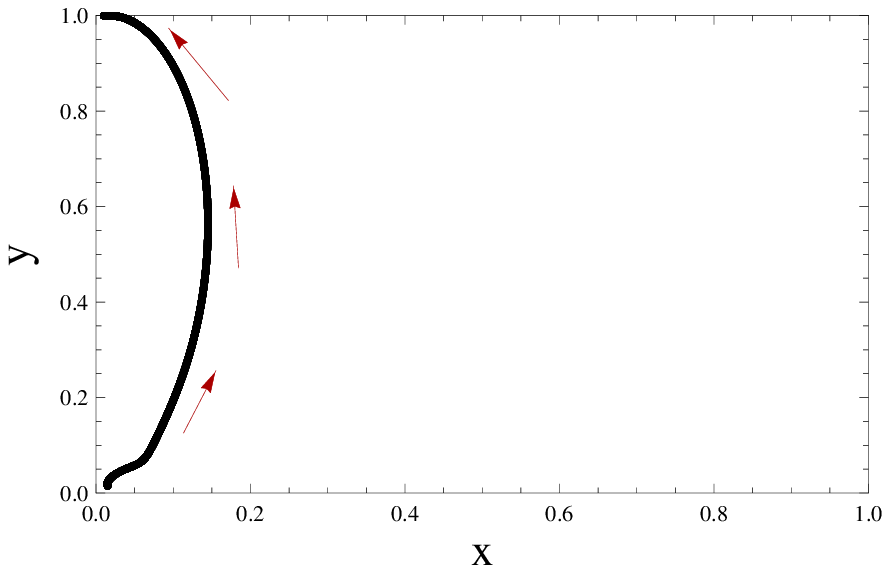}\label{fig3}}
\subfigure[]{\includegraphics[width=60mm, height=54mm]{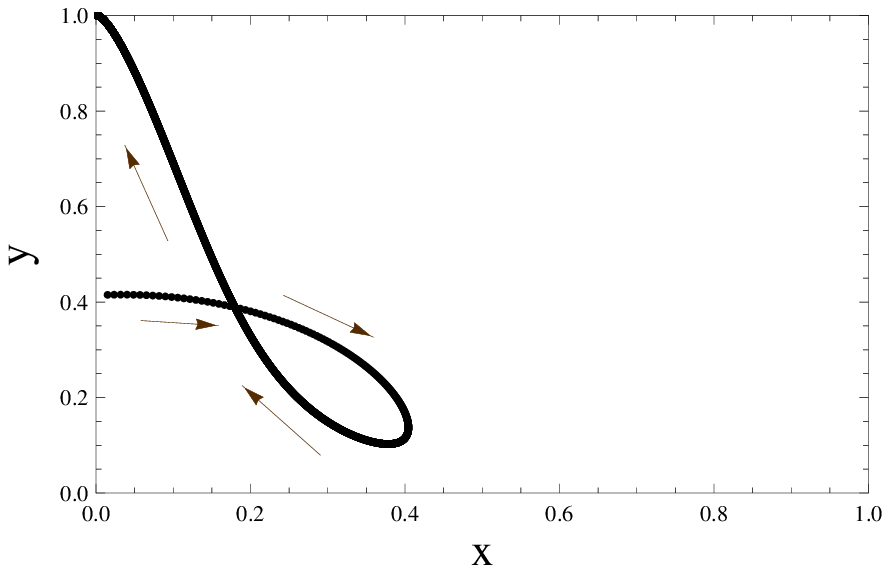}\label{fig4}}
\subfigure[]{\includegraphics[width=60mm, height=54mm]{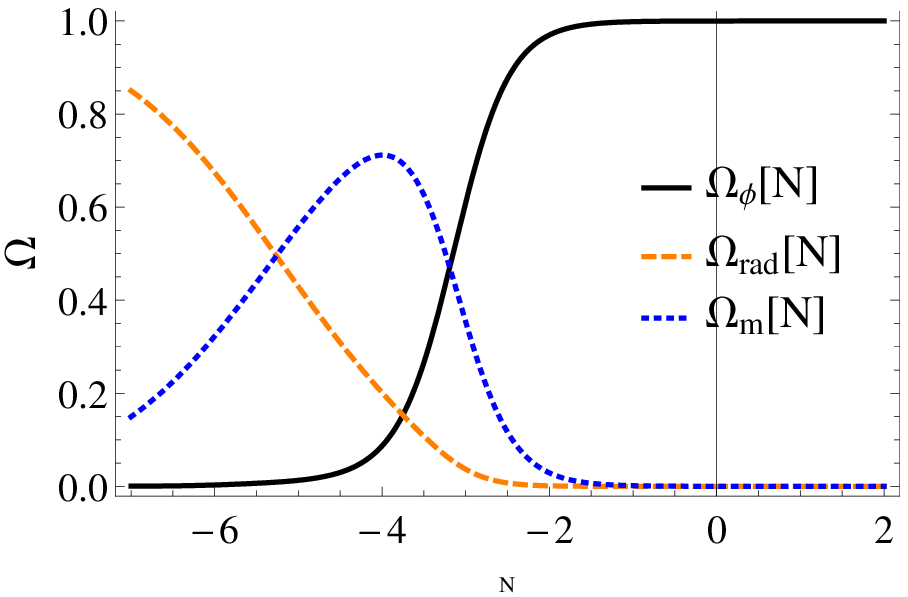}\label{fig1}}
\subfigure[]{\includegraphics[width=60mm, height=54mm]{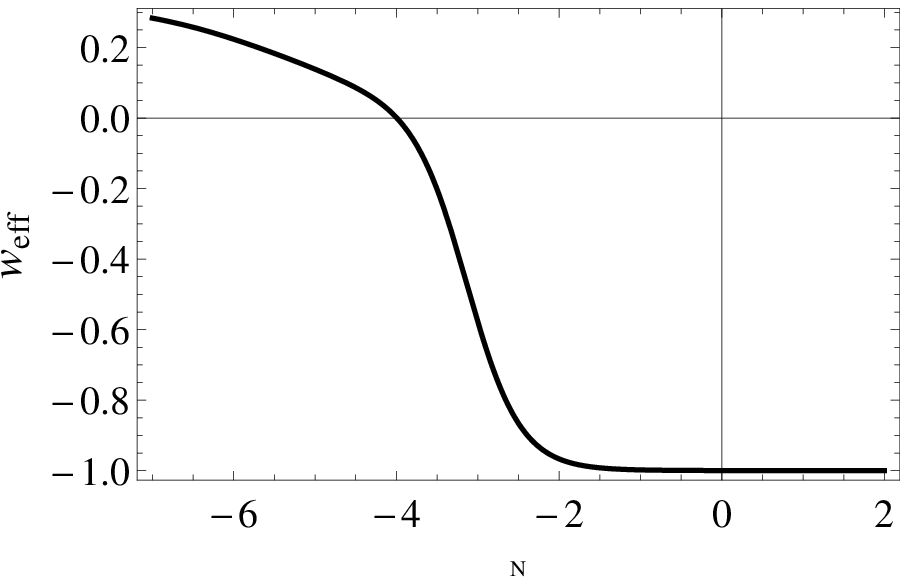}\label{fig2}}
\caption{ (Color online). Dynamics of scalar field in the phase space when there is no interaction for
a double exponential potential for different initial conditions. Convergence of both curves is at the same point. (c) Evolution of different energy densities $\Omega_i$ w.r.t. $N$ and (d) evolution of $w_{eff}$ w.r.t. $N$. The initial conditions for the preceding evolution were $x_i=0.015$, $y_i=0.015$, $u_i=\sqrt{0.85}$ and $z_i=40$, and $\lambda=6$.}\label{fig:b0}
\end{figure}

\begin{figure}[t!]
\centering
\subfigure[]{\includegraphics[width=60mm, height=54mm]{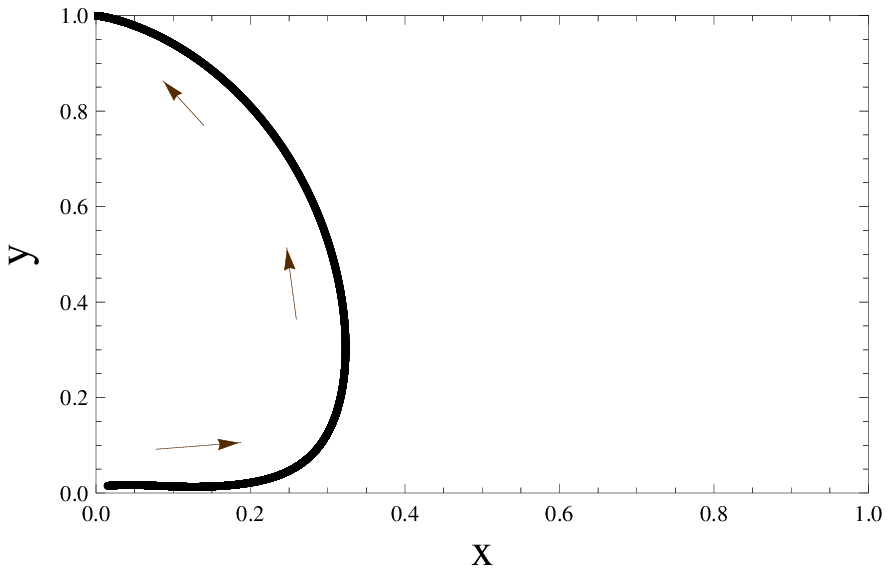}\label{psic1}}
\subfigure[]{\includegraphics[width=60mm, height=54mm]{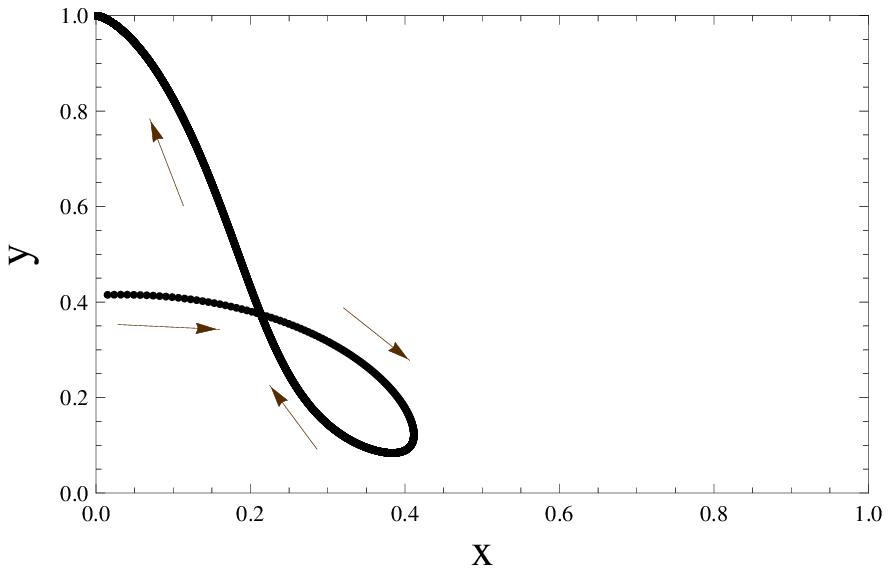}\label{psic2}}
\subfigure[]{\includegraphics[width=60mm, height=54mm]{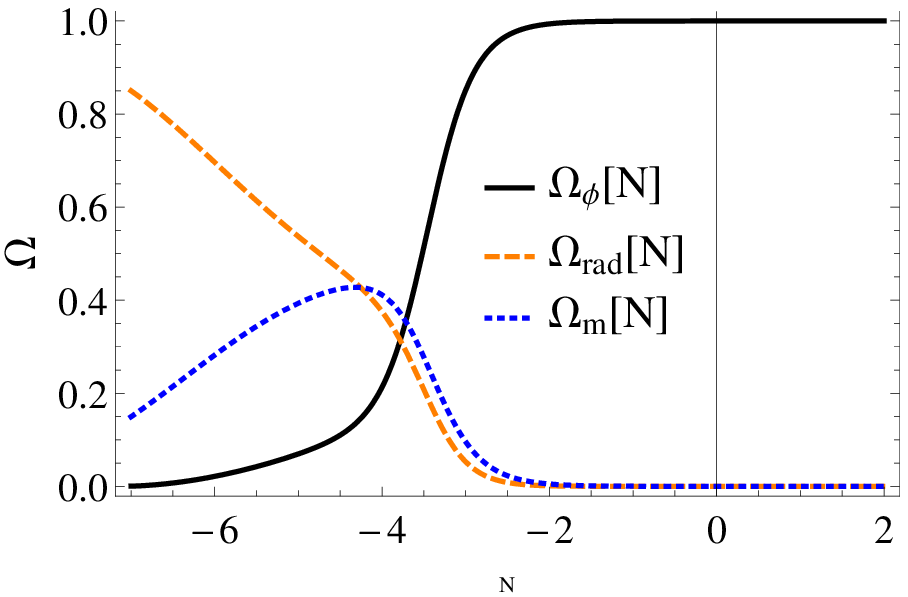}\label{omegas}}
\subfigure[]{\includegraphics[width=60mm, height=54mm]{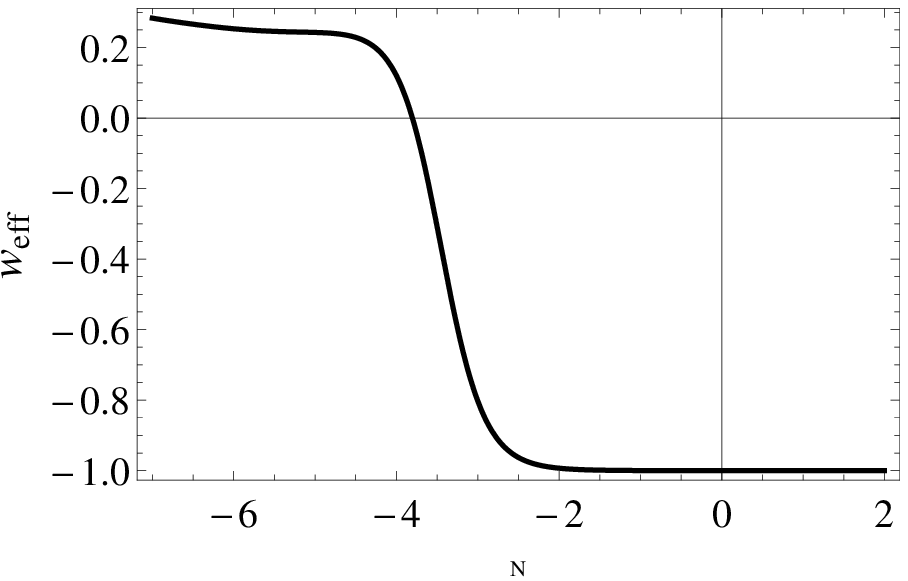}\label{w-e}}
\caption{ (Color online). Dynamics of scalar field in the phase space when interaction is switched on for
a double exponential potential for the same  initial conditions as in figure~\ref{fig:b0}. Convergence of both curves is at the same point. Also as in figure~\ref{fig:b0}, the evolution of different energy densities (c) $\Omega_i$,  and (d) $w_{eff}$ w.r.t. $N$, respectively.}\label{fig:b1}
\end{figure}

The dynamical system framework~\eqref{dyn-sys} introduced so far is general enough to accommodate analysis for large class of potentials with or without any interaction. Furthermore, the nature of the field (whether quintessence or phantom) has been properly incorporated into the framework.  
  In the next section we shall work with a specific kind of potential and interaction term. Once we know the form of interaction $Q_I$ and the potential $V(\phi)$, we can obtain the critical points $(x_*,y_*,u_*,z_*)$ of the autonomous system~\eqref{dyn-sys} by imposing the conditions $x_*^\prime=y_*^\prime=u_*^\prime=z_*^\prime=0$. These critical points must be real to ensure their existence in the phase space. To study the stability of the critical points, let us consider small perturbations $\delta x,\delta y,\delta u$ and $\delta z$ around the critical points $(x_*,y_*,u_*,z_*)$, i.e.,
\begin{equation}\label{perturb-var}
  x\rightarrow x_*+\delta x,\quad y\rightarrow y_*+\delta y,\quad u\rightarrow u_*+\delta u,\quad z\rightarrow z_*+\delta z \,.
\end{equation}
Combining the four variables into a vector $\bf x_a$, we have  $\bf x_a\rightarrow x_{a*
}+\delta x_a$. On substituting perturbed variables from~\eqref{perturb-var} into the system~\eqref{dyn-sys}, we get a set of first order differential equations symbolically represented by
\begin{equation}
\frac{d}{dN}(\delta \bf x_a\\)\rm=M_{\bf x_{a*}}(\delta \bf x_a\\)  \,,
\end{equation}
where $ \delta \bf x_a$ and $M_{\bf x_{a*}}$ are the column vector of the perturbations and the coefficient matrix, respectively. $M_{\bf x_{a*}}$ depends on the critical points and is represented as
\[ M_{\bf x_{a*}}\equiv \left(\begin{array}{cccc}
\frac{\partial x^\prime}{\partial x} & \frac{\partial x^\prime}{\partial y}& \frac{\partial x^\prime}{\partial u} & \frac{\partial x^\prime}{\partial z}\\
\frac{\partial y^\prime}{\partial x} & \frac{\partial y^\prime}{\partial y} & \frac{\partial y^\prime}{\partial u} &\frac{\partial y^\prime}{\partial z}\\ \frac{\partial u^\prime}{\partial x} & \frac{\partial u^\prime}{\partial y} & \frac{\partial u^\prime}{\partial u} & \frac{\partial u^\prime}{\partial z}\\ \frac{\partial z^\prime}{\partial x} & \frac{\partial z^\prime}{\partial y} & \frac{\partial z^\prime}{\partial u} & \frac{\partial z^\prime}{\partial z}  
\end{array} \right)\\.   \]
The nature of the four eigenvalues of the coefficient matrix$M_{\bf x_{a*}}$ determines the stability of the critical points. The criteria for establishing the stability of the critical points is as follows:
\\$a)$S
node   : if all the eigenvalues are negative.
\\$b)$Unstable node : if all the eigenvalues are positive.
\\$c)$Saddle point  : if some eigenvalues are positive and some are negative.
\\\\However, if any one or more eigenvalues vanish, the linear stability analysis  breaks down. In that case, the center manifold theory needs to be applied~\cite{coley03, wainwright05}. The stability for the special case can also be determined from the signs of other non-null eigenvalues if the critical points belong to normally hyperbolic sets~\cite{leon14}.

\section {Case of double exponential potential}

Upto now, we have not considered any specific form of $V$ and $Q_I$. Now onwards we shall work with $\epsilon =1$ (quintessence) for the sake of simplicity. Moreover, we shall assume that $Q_I$ is $\sqrt{\frac{2}{3}}\kappa \beta \dot \phi\rho_m$ which is also a simple and widely studied interaction in a variety of potentials~\cite{copeland06,chen09,boehmer08}.

We choose the double exponential potential for our quintessence field. Although this potential is significant from the viewpoint of fundamental theories, particularly of higher dimensions~\cite{huey00}, it is still under-studied potential in the literature for dark-energy modeling.  It is to be mentioned that this potential is different from the sum of two exponential terms~\cite{leon14} which is also known by the same name.  

On putting potential
\begin{equation}\label{potential}
V(\phi)=V_0\exp{(e^{-\kappa \lambda \phi})} 
\end{equation}
\\in Eqs.~\eqref{def:third-var}-\eqref{def:gamma}, we get
\begin{equation}
z=\lambda \exp^{-\kappa \lambda \phi} \,,
\end{equation}
\begin{equation}\label{}
\Gamma-1=\frac{\lambda}{z}
\end{equation}
Also, using $Q_I=\sqrt{\frac{2}{3}}\kappa\beta\dot\phi\rho_m$ in eq.~\eqref{def:interaction}, we get
\begin{equation}
Q=\beta\ (1-x^2-y^2-u^2) 
\end{equation}
\\Therefore, the dynamical system~\eqref{dyn-sys} transforms as
\begin{align}
x^\prime &= -3x+\sqrt{\frac{3}{2}}y^2z+\frac{3}{2}x[(1-x^2-y^2-u^2)+\frac{4}{3}u^2+2x^2]-\beta\ (1-x^2-y^2-u^2) \\
y^\prime &=-\sqrt{\frac{3}{2}}yxz+\frac{3}{2}y[(1-x^2-y^2-u^2)+\frac{4}{3}u^2+2x^2] \\
u^\prime &=-2u+\frac{3}{2}u[(1-x^2-y^2-u^2)+\frac{4}{3}u^2+2x^2] \\
z^\prime &=-\sqrt{6}zx\lambda
\end{align}
Also, from Eqs.~\eqref{friedmann-constraint} and ~\eqref{raychaudhuri}, we have 
\begin{equation}
\frac{\dot H}{H^2}=-\frac{3}{2}[1+w_{eff}] \label{w-H}
\end{equation}
where $w_{eff}$ in terms of dimensionless variables~\eqref{def:var} is 
\begin{equation}\label{w-eff}
w_{eff}=2x^2-1+\frac{4}{3}u^2+(1-x^2-y^2-u^2) \,. 
\end{equation}

To have acceleration, we need $w_{eff}<-\frac{1}{3}$.

\begin{table}
\centering
\caption{Critical points of the system and their relevant properties for double exponential potential. \label{table:cri-pts}}
\centering
\begin{tabular}{||l|c|c|c|c|c|c|c|c||}
\hline
\hline
pt. & $x_c$ &$y_c$ &$u_c$ &$z_c$ &$\Omega_\phi$ &$\gamma_\phi$ &$\omega_\phi$&Existence
\\
\hline
\hline
A&0&0&1& z&0 &Indet. &Indet.& $\forall \beta$
\\[2ex]
B&-1&0&0&0&1 &2 &1 & $\forall \beta$
\\[2ex]
C&0&0&1&0 &0 & Indet.&Indet. & $\forall \beta$
\\[2ex]
D&1&0&0&0&1 &2 &1 & $\forall \beta$
\\[2ex]
E&$\frac{1}{2\beta}$&0&$\frac{\sqrt{-3+4\beta^2}}{2\beta}$&0 & $\frac{1}{4\beta^2}$&2 & 1&$\beta^2>\frac{3}{4}$
\\[2ex]
F &$\frac{2\beta}{3}$& 0& 0&0&$\frac{4\beta^2}{9}$ & 2& 1& $\forall \beta$
\\[2ex]
G&0&1&0&0& 1&0 & -1& $\forall \beta$
\\[2ex]
H&$-\frac{3}{2\beta}$&$\frac{\sqrt{4\beta^2+9}}{2\beta}$&0&0&1+ $\frac{9}{2\beta^2}$&$\frac{9}{2\beta^2+9}$ & $\frac{-2\beta^2}{2\beta^2+9}$&$\beta^2>\frac{-9}{4}$
\\[2ex]
\hline
\end{tabular}
\end{table}

\begin{table}
\centering
\caption{Eigenvalues for the critical points, their stability properties and energy densities of various components. See text for $a, b, c$, $\mu1$ and $\mu2$.\label{table:cri-pts-prop}}
\centering
\begin{tabular}{|l|c|c|c|c|c|c|r|}
\hline
\hline
$pt.$ & $w_{eff}$ &q&$\frac{\Omega_\phi}{1-\Omega_\phi}$&E.V's&$\Omega_r$&$\Omega_m$&Stability
\\
\hline
\hline
A&$\frac{1}{3}$&1&Indet.&$(2,-1,0,1)$& 1&0&Saddle
\\[2ex]
B& 1&2&Indet.&$(3,1,3+2\beta,\sqrt{6}\lambda)$& 0&0&Unstable
\\[2ex]
C&$\frac{1}{3}$&1&Indet.&$(2,-1,0,1)$& 1&0&Saddle
\\[2ex]
D&1&2&Indet.&$(3,1,3-2\beta,-\sqrt{6}\lambda)$& 0&0&Saddle
\\[2ex]
E&$\frac{1}{3}$&1&$\frac{1}{2}$&$(\mu1,\mu2,2,-\sqrt{\frac{3}{2}}\frac{\lambda}{\beta})$& $1-\frac{3}{4\beta^2}$ &$\frac{1}{2\beta^2}$&Saddle
\\[2ex]
F&$\frac{4\beta^2}{9}$& $\frac{3+4\beta^2}{6}$& $\frac{4\beta^2}{-4\beta^2+9}$&$(a,b,c,-2\sqrt{\frac{2}{3}}\beta\lambda)$& 0&$1-\frac{4\beta^2}{9}$ &Saddle
\\[2ex]
G&-1&-1&Indet.&$(-3,-2,0,-3)$&0&0 &stable
\\[2ex]
H&-1&-1&$-1-\frac{2\beta^2}{9}$&$(\mu3,\mu4,-2,\frac{3\sqrt{\frac{3}{2}}\lambda}{\beta})$& 0&$\frac{-9}{2\beta^2}$& Saddle
\\[2ex]
\hline
\hline
\end{tabular}
\end{table}

The auxiliary variables $a, b, c$, $\mu1$, $\mu2$, $\mu3$ and $\mu4$ in table~\ref{table:cri-pts-prop} are defined as

\begin{align}
\mu1 &\equiv\frac{-1}{2\beta^2}(\beta^2+\sqrt{3}\sqrt{\beta^2-\beta^4}) \,,\\
\mu2 &\equiv\frac{1}{2\beta^2}(-\beta^2+\sqrt{3}\sqrt{\beta^2-\beta^4}) \,,\\
\mu3 &\equiv\frac{-3}{2\beta^2}[\beta^2+\sqrt{\beta^2(9+5\beta^2)} ] \,,\\
\mu4 &\equiv\frac{3}{2\beta^2}[-\beta^2+\sqrt{\beta^2(9+5\beta^2)} ] \,,\\
a &\equiv\frac{4\beta^2+9}{6} \,,\\
b &\equiv\frac{4\beta^2-9}{6} \,,\\
c &\equiv\frac{4\beta^2-3}{6} \,.
\end{align}
\\\\ Now we shall scrutinize each critical point carefully:
\\\\
$a)$ From table~\ref{table:cri-pts}, it is clear that the constraint $\beta^2>\frac{-9}{4}$ for the existence of point H can not be satisfied with the constraint equation which needs $\frac{18}{4\beta^2}\leq 0$
for any real $\beta$. Therefore this point is excluded from further consideration.
\\$b)$ As $w_{eff}=\frac{1}{3}$ and $q=1$ for points A, C and E (see table~\ref{table:cri-pts-prop}), the scalar field mimics the nature of radiation while the universe is decelerating for these points. As these are the saddle points too, this is just a transient behavior.
\\c)  As $\Omega_\phi=1$ and $q=1$ for unstable points B and D (see table~\ref{table:cri-pts-prop}), these are the dark energy dominated solutions with the universe decelerating.

 d) The only attractor we have in our case is the point G. This point has $w_{eff}=-1$ (from figure~\ref {fig2}) and $q=-1$, which are the conditions to be satisfied for accelerated expansion (see eq.~\eqref{w-eff}). On putting $w_{eff}=-1$ in eq.~\eqref{w-H}, $\dot H=0$ vanishes, in other words, $H$ becomes constant, which leads to $a \propto e^{Ht}$. This implies that we have a de-Sitter solution at the present epoch. No scaling solution is observed for this point since $\Omega_\phi=1$. However, the interesting aspect of this point is that we have a global attractor at the present epoch. It means that  our trajectories converge at this point independent of the initial conditions (see figure~\ref{fig3}and figure~\ref {fig4}).

 From figure~\ref{fig1}, it is apparent that when there is no interaction (i.e., $\beta=0$), the above defined autonomous system follows the desired cosmological dynamics. It describes an early radiation-dominated era(RDE, dominated by $\Omega_{rad}$) followed by a matter dominated epoch (MDE,dominated by $\Omega_m$), and finally the universe enters into the present accelerated regime (dominated by $\Omega_\phi$).

\section{Conclusion and outlook}
In this work, the deSitter solution is obtained at the present epoch to account for the accelerated regime. Dynamical evolution of the field translates into a smooth transition of the universe from a decelerated epoch to an accelerated one in the presence of a dark energy component --- a scalar field having a self interaction described in terms of a double exponential potential. Suitable combinations of initial conditions and parameter values are found for which a realistic sequential dominance of different cosmological constituents could follow. It is found that the initial dynamical evolution is dependent on the interaction parameter $\beta$; however, the asymptotic behavior of the system is independent of the value of $\beta$ --- typical of an attractor solution(see figure \ref{fig1} and \ref{omegas}). It is so because there exists only one stable attractor, i.e., point G. As the point G admits a deSitter solution (complete DE dominated), it is not possible to address the coincidence problem strictly within the scenario considered.

As  alternative direction for future work, it would be interesting to explore if  other possibilities such as time-dependent interaction,  non-minimal coupling and phantom field (usually considered in the literature~\cite{ copeland06, boehmer08, chen09}), might provide a global attractor for which $\Omega_i$s are parameter dependent and which could account for the coincidence problem. 

\section{Acknowledgement}
We are grateful to several members of Department of Physics and Astrophysics for their useful comments. We would also like to thank Shiv Sethi for a critical review of the manuscript, and Sanjeev Kumar for help in numerical methods. VG wishes to thank the CSIR (India) for the financial support through grant number 09/045/(0933)/2010 EMR-I.

\bibliographystyle{JHEP}
\bibliography{quint}

\end{document}